\documentclass[floatfix,aps,prd,twocolumn,showpacs,nofootinbib]{revtex4} 

\usepackage{graphicx}
\usepackage{amsmath}
\usepackage{amssymb}

\begin{document}



\title{Chaos, Determinacy and Fractals in Active-Sterile Neutrino
  Oscillations\\ in the Early Universe}

\author{Kevork N. Abazajian}

\email{kev@umd.edu}

\affiliation{Department of Physics, University of Maryland, College
  Park, MD 20742-4111}

\author{Prateek Agrawal}

\email{apr@umd.edu}

\affiliation{Department of Physics, University of Maryland, College
  Park, MD 20742-4111}

\begin{abstract}
  The possibility of light sterile neutrinos allows for the resonant
  production of lepton number in the early universe through
  matter-affected neutrino mixing.  For a given a mixing of the active
  and sterile neutrino states it has been found that the lepton number
  generation process is chaotic and strongly oscillatory.  We
  undertake a new study of this process' sensitivity to initial
  conditions through the quantum rate equations.  We confirm the
  chaoticity of the process in this solution, and moreover find that
  the resultant lepton number and the sign of the asymmetry produces a
  fractal in the parameter space of mass, mixing angle and initial
  baryon number.  This has implications for future searches for
  sterile neutrinos, where arbitrary high sensitivity could not be
  determinate in forecasting the lepton number of the universe.
\end{abstract}
\maketitle

\section{introduction}

Neutrino physics is entering an era of increasing precision as initial
evidence for neutrino oscillations is leading to confirmation and
refinement of the neutrino mass and mixing parameters.  The solar
neutrino deficit is consistent with a matter-affected
Mikheyev-Smirnov-Wolfenstein (MSW) transformation of electron
neutrinos, $\nu_e$~\cite{Ahmed:2003kj}, and was confirmed in the CPT
conjugate channel by reactor electron antineutrino disappearance at
KamLAND~\cite{Araki:2004mb}.  Atmospheric neutrino disappearance due
to apparent neutrino oscillations between what is predominantly
$\nu_\mu$ and $\nu_\tau$ and their CPT conjugates has been confirmed
and is being determined precisely by long-baseline neutrino
experiments ~\cite{Michael:2006rx,Ahn:2006zza}.

There were also indications of neutrino oscillations governed by a
large mass scale from the short-baseline Liquid Scintillator Neutrino
Detector (LSND)~\cite{Athanassopoulos:1997pv} that would require a
third mass scale difference and therefore a fourth neutrino or
more~\cite{Sorel:2003hf}.  This neutrino could not participate in weak
interactions due to the $Z^0$ width~\cite{Yao:2006px}, and therefore
is dubbed ``sterile.''  Sterile neutrinos are prevalent in most models
of neutrino mass generation mechanisms~\cite{Mohapatra:2006gs}, and
could be light enough to participate in neutrino oscillations.  The
four neutrino oscillation hypothesis for the LSND result is in
conflict with new results from the from the short-baseline MiniBOONE
experiment\cite{collaboration-2007-98}.  The parameter space of
interest for the LSND and MiniBOONE results are a small fraction of
that of interest for lepton number generation, and do not exclude the
possibility of this mechanism. Moreover, such light sterile neutrinos
in the parameter space of interest for lepton number generation, and
outside of the MiniBooNE bounds are favored to produce the electron
fraction required by successful $r$-process nucleosynthesis in Type-II
supernovae \cite{McLaughlin:1999pd,Fetter:2002xx}.  See-saw type mass
generation mechanisms could readily include such light sterile
neutrinos \cite{deGouvea:2005er}.

The effects of sterile neutrinos in the early universe have been under
study for some time.  Early work studied the effect of the partial or
complete thermalization of the sterile neutrino through oscillations,
which would alter the energy density and expansion rate of the
universe prior to primordial nucleosynthesis, and therefore the
production of the light
elements~\cite{Langacker:1989sv,Barbieri:1989ti,Kainulainen:1990ds,Shi:1993hm}.
The primordial helium yield is very sensitive to the expansion rate,
and has been used to place bounds on four neutrino models with
appreciable mixing between active and sterile neutrinos of the kind
required for the LSND
indications~\cite{DiBari:2001ua,Abazajian:2002bj}.  It was also
discovered that an MSW resonance can cause the generation of a large
neutrino asymmetry that could modify primordial nucleosynthesis even
with very small mixing angles between the active and sterile
neutrinos~\cite{Foot:1995qk}.  This was confirmed in
Ref.~\cite{Shi:1996ic}, and further found rapid oscillations in the
asymmetry generation, extreme sensitivity to initial conditions, and
chaoticity in certain regions of parameter space of mass and mixing
angle.

A key question regarding the chaoticity of the lepton generation
mechanism is the determinacy of the sign of the asymmetry given a
mixing angle and mass-splitting for the active-sterile neutrino pair.
That is, if a mass and mixing angle were experimentally measured for
the pair, could the sign and magnitude of the cosmological neutrino
asymmetry be predicted?  One way of assessing the predictability of
the final lepton number given a potentially experimentally determined
mass and mixing angle is the level of information loss in the lepton
number generation process.  The level of information loss in the
chaotic process was explored in Ref.~\cite{Braad:2000zw}, which found
appreciable loss of initial conditions due to chaoticity in small
regions of parameter space.  The sensitivity of the sign and amplitude
of the asymmetry to the mixing parameter space was found to be
``complex'' in Ref.~\cite{enqvist-1999-464}.  

All of the work discussed thus far was done through the quantum rate
equations (QRE's), where the neutrino system is modeled to follow the
average neutrino momentum $\langle p \rangle \approx 3.15 T$.  The
strong momentum dependence of the resonant process of lepton number
generation suggests that an accurate solution should take into account
the full spectrum in the solution of the equations through the
so-called Quantum Kinetic Equations (QKE's).  The narrowness of the
resonance requires exceptionally fine momentum binning and therefore
following the exponential growth in a very large number of momenta
modes.  Several authors found limitations to solving the equations
numerically \cite{bari-foot,Volkas:2000ei}.  For example,
Ref.~\cite{bari-foot} found oscillatory behavior in a range of
parameter space with the QKE's, but could not conclude that it was
physical and not numerical.  The QKE's were solved through the
oscillatory phase by Ref.~\cite{kainulainen-2002-0202}, who used a new
binning procedure following the resonance to simulaneously have fine
binning yet a manageable number of momenta modes. They found that the
chaotic nature of the generation process remained.

In this work, we explore the nature of the lepton number generation
process and its chaoticity.  We confirm that the resonant generation
is highly oscillatory in regions of the parameter space and extremely
sensitive to the initial conditions.  Moreover, we explore the
dependence of the final sign of the asymmetry on the oscillation
parameter space and find that the extreme sensitivity to the parameter
space gives rise to a fractal structure in the final sign dependence
on mass-splitting and mixing angle.  This has significant implications
for the determinacy of the cosmological lepton number in the case of
discovering a small sterile neutrino mixing with the active neutrino
sector.  If the parameters are found to lie in a region of fractal
nature, even arbitrarily high precision in the determination of the
parameters will not allow for a prediction of the cosmological lepton
number.

%
%

We will briefly introduce our system in Sec. \ref{sec:QKEs}, including
the effects of our constant momentum approximation. We go on in
Sec. \ref{sec:results} to present the solution under the above
approximation, where we show that the final sign of the asymmetry is
not just chaotic, but exhibits a self-similarity like that of a
fractal at very small scales.  We compare our results with
Ref. \cite{enqvist-1999-464}, which shows that the resolution and
numerical error are responsible for the seeming ``white noise''
structures in the parameter space.  We conclude in
Sec. \ref{sec:Conclusion}.

\section{The Quantum Kinetic and Rate Equations\label{sec:QKEs}}

For specificity, we concentrate on oscillations between the
$\tau$-neutrino ($\nu_{\tau}$) and a sterile neurino ($\nu_{s}$).  The
vacuum mixing angle is defined by the following equation,
\begin{equation}
\left[\begin{array}{c}
\nu_{\tau}\\
\nu_{s}\end{array}\right]=\left[\begin{array}{cc}
\cos\left(\theta_{0}\right) & \sin\left(\theta_{0}\right)\\
-\sin\left(\theta_{0}\right) & \cos\left(\theta_{0}\right)
\end{array}\right]\left[
\begin{array}{c}
\nu_{a}\\
\nu_{b}\end{array}\right]
\end{equation}
where $\nu_{a}$ and $\nu_{b}$ are mass eigenstates. Conventionally,
$\cos(\theta)$ is chosen to be positive, and $\delta
m^{2}\equiv m_{b}^{2}-m_{a}^{2}$.

An $\alpha$-type neutrino asymmetry is defined as 
\begin{equation}
L_{\nu_{\alpha}}\equiv\frac{n_{\nu_{\alpha}}-n_{\bar{\nu}_{\alpha}}}{n_{\gamma}},
\end{equation}
where the photon number density $n_{\gamma}=2\zeta(3)T^{3}/\pi^{2}$.
The effective total lepton number $L^{(\alpha)}$, useful when defining
the evolution equations below, is given by
\begin{eqnarray}
L^{(\alpha)} & = &
L_{\nu_{\alpha}}+L_{\nu_{e}}+L_{\nu_{\mu}}+L_{\nu_{\tau}}+\eta
\end{eqnarray}
where $\eta\equiv n_{\rm b}/n_\gamma$ is the baryon to photon ratio at the epoch of oscillations.

The density matrix parametrization for an active-sterile oscillation
can be chosen as
\begin{eqnarray}
  \rho_{\tau\beta'}(p) & = & \frac{1}{2}[P_{0}(p)I+\mathbf{P}(p). \mathbf{\sigma}]\nonumber \\
  \bar{\rho}_{\tau\beta'}(p) & = &
  \frac{1}{2}[\bar{P}_{0}(p)I+\bar{\mathbf{P}}(p).\mathbf{\sigma}],
\end{eqnarray}
where $\mathbf{P}(p) =
P_{x}(p)\hat{\mathbf{x}}+P_{y}\hat{\mathbf{y}}+P_{z}\hat{\mathbf{z}}$
is the polarization vector;
$\mathbf{\sigma}=\sigma_{x}\hat{\mathbf{x}} +
\sigma_{y}\hat{\mathbf{y}}+\sigma_{z}\hat{\mathbf{z}}$ are the usual
Pauli matrices. The evolution of $\mathbf{P}(p)$, $P_{0}(p)$,
$\bar{\mathbf{P}}(p)$ and $\bar{P}_{0}(p)$ is given by the following
equations \cite{kainulainen-2002-0202},
\begin{eqnarray}
  \frac{d\mathbf{P}}{dt} & = & \mathbf{V}(p) \times \mathbf{P}(p) - 
     D(p)[P_{x}(p)\hat{\mathbf{x}} + 
     P_{y}(p)]\hat{\mathbf{y}}] + 
     \frac{d P_{0}}{dt}\hat{\mathbf{z}}\nonumber \\
  \frac{dP_{0}}{dt} & \simeq &
  \Gamma(p)[\frac{f_{eq}(p)}{f_{0}(p)}-\frac{1}{2}(P_{0}(p)+P_{z}(p))].
\label{eq:qke}
\end{eqnarray}

The equations for the anti-particles are given by the obvious
substitutions, $\mathbf{P} \rightarrow \bar{\mathbf{P}}$,
$P_{0}\rightarrow\bar{P}_{0}$,
$\mathbf{V}(p)\rightarrow\bar{\mathbf{V}}(p)$, $f_{eq}(p) \rightarrow
\bar{f}_{eq}(p)$.  $\bar{\mathbf{V}}(p)$ is obtained by replacing
$L^{(\tau)}$by $-L^{(\tau)}$. The total collision rates of the flavor
neutrino and anti-neutrino are approximately equal,
$\Gamma(p)\simeq\bar{\Gamma}(p)$.  These 8 equations form the quantum
kinetic equations for active-sterile oscillation.

The damping coefficient is given by
\begin{equation}
D(p)=\Gamma(p)/2,
\end{equation}
where $\Gamma(p)$ is the total collision rate of the neutrino, given
by $0.92G_{F}^{2}T^{5}p$ for the $\tau$-neutrino. Further, $\bar{D}\simeq D$
to a good approximation \cite{enqvist-1999-464}.

The rotation vector $\mathbf{V}(p)$ has the following components
\begin{eqnarray}
  V_{x}(p) & = & \frac{\delta m^{2}}{2p}\sin2\theta_{0}\nonumber \\
  V_{y}(p) & = & 0\nonumber \\
  V_{z}(p) & = & V_{0}(p)+V_{L}(p),
\end{eqnarray}
where
\begin{eqnarray}
  V_{0}(p)&=&-\frac{\delta m^{2}}{2p}\cos2\theta_{0}+V_{1}\nonumber \\
  V_{1}(p)&=&-\frac{7\sqrt{2}}{2}\frac{\zeta(4)}{\zeta(3)} \frac{G_{F}}{M_{z}^{2}}n_{\gamma}pT[n_{\nu_{\tau}}+n_{\nu_{\bar{\tau}}}]\nonumber \\
  V_{L}(p)&=&\sqrt{2}G_{F}n_{\gamma}L^{(\tau)}.\nonumber
\end{eqnarray}
Here $n_{\gamma}$ is the photon equilibrium number density,
$n_{\nu_{\tau}}\text{ and }n_{\nu_{\bar{\tau}}}$ are normalized to
unity and $f_{eq}(p)$ is the Fermi-Dirac distribution with a chemical
potential $\mu_{\tau}$.

The chaotic nature of the solution is seen to not depend on
$\eta$, which was set to $\sim 1\times10^{-10}$ (though the sign
of the asymmetry can be sensitive to this initial value as well). The
well-known Fermi-Dirac distributions are
\begin{eqnarray}
f_{0}(p) & = & \frac{1}{1+\exp(p/T)}\nonumber \\
f_{eq}(p) & = & \frac{1}{1+\exp(p/T-\tilde{\mu}_{\tau})},
\end{eqnarray}
where $\tilde{\mu}_{\tau}\equiv\mu_{\nu_{\tau}}/T$. To get the
distribution for anti-neutrinos, we substitute
$\tilde{\mu}_{\tau}\rightarrow\tilde{\mu}_{\bar{\tau}}$.  The chemical
potentials can be found using the neutrino asymmetry.  For a system in
thermal equilibrium, the asymmetry is given by
\begin{equation}
L_{\nu_{\tau}}=\frac{1}{4\zeta(3)}\int_{0}^{\infty}
\frac{x^{2}dx}{1+e^{x-\tilde{\mu}_{\tau}}} - 
\frac{1}{4\zeta(3)}
\int_{0}^{\infty}\frac{x^{2}dx}{1+e^{x-\tilde{\mu}_{\bar{\tau}}}}.
\end{equation}
We can expand the above integral in
$\left(\tilde{\mu}_{\tau}-\tilde{\mu}_{\bar{\tau}}\right)$ 
\begin{eqnarray}
  L_{\nu_{\tau}} &\simeq& \frac{1}{24\zeta(3)}
  \Big[\pi^{2}\left(\tilde{\mu}_{\tau} -
      \tilde{\mu}_{\bar{\tau}}\right)\nonumber \\ 
   && +\ 6\left(\tilde{\mu}_{\tau}^{2}
      - \tilde{\mu}_{\bar{\tau}}^{2}\right) \ln(2) +
    \left(\tilde{\mu}_{\tau}^{3} - \tilde{\mu}_{\bar{\tau}}^{3}\right)\Big].
\end{eqnarray}
We need another relation between $\tilde{\mu}_{\tau}$ and
$\tilde{\mu}_{\bar{\tau}}$ to solve for them. For temperatures greater
than the chemical decoupling temperatures, pair creation and
annihilation processes are fast enough to ensure chemical
equilibrium. However, below these temperatures, neutrinos no longer
remain in thermal equilibrium, though weak interactions are fast
enough to thermalize them. As a consequence, the value of
$\tilde{\mu}_{\tau}$ remains fixed at the decoupling temperature, but
$\tilde{\mu}_{\bar{\tau}}$ continues to decrease.  In summary
\begin{eqnarray}
  \tilde{\mu}_{\tau}+\tilde{\mu}_{\bar{\tau}}=0 
&  & T\gtrsim T_{dec}^{\tau}\nonumber \\
  \tilde{\mu}_{\tau}=\tilde{\mu}_{\tau}(T=T_{dec}^{\tau}) &  & T\lesssim
  T_{dec}^{\tau},
\end{eqnarray}
where $T_{dec}^{\tau}\simeq 3.5$ MeV. These two equations above determine
the chemical potentials to be used in the Fermi-Dirac distributions.

We employ the approximation that all the neutrinos have the same
momentum, {\it viz.} $p = \langle p \rangle \approx 3.15 T$ for our
analysis. The approximation inflates the rate at which neutrino
asymmetry is created and therefore may be unsuitable for the thermal
regions where the asymmetry growth is shown to be exponential
\cite{bari-foot}. Apart from this shortcoming, this approximation has
the significant advantage of being numerically tenable, and is useful to
explore the general features of the system.  We also carried out
analysis using the full momentum dependent quantum kinetic equations
(QKEs). The full system is too numerically intensive to run over the
entire parameter range at high resolution to study the fractal
structure. Moreover, our work with the full QKEs never reached the
numerical stability which we required, and describe below.

Further, following the work of Ref. \cite{enqvist-1999-464}, we
approximate $P_{0}$ to be a constant, set to $1$. The asymmetry is
given simply by
\begin{equation}
L_{\nu_{\tau}}=\frac{3}{8}\left(P_{z}-\bar{P_{z}}\right).
\end{equation}

In order to achieve numerical efficiency and stability, we rewrite
$\{P_{i},\bar{P}_{i}\}$ in the above equation as
$\{P_{i}^{+},P_{i}^{-}\}$, where $P_{i}^{\pm}=P_{i}\pm\bar{P_{i}}$.
This separates out the large quantities and the small quantities in
the differential equation \cite{enqvist-1999-464}. Thus, the equations
take the form
\begin{eqnarray}
\dot{P}_{x}^{+} & = & -DP_{x}^{+}-V_{0}P_{y}^{+}-V_{L}P_{y}^{-}\nonumber \\
\dot{P}_{y}^{+} & = & -DP_{y}^{+}-V_{x}P_{z}^{+}+V_{0}P_{x}^{+}+V_{L}P_{x}^{-}\nonumber \\
\dot{P}_{z}^{+} & = & V_{x}P_{y}^{+}\\
\dot{P}_{x}^{-} & = & -DP_{x}^{-}-V_{0}P_{y}^{-}-V_{L}P_{y}^{+}\nonumber \\
\dot{P}_{y}^{-} & = & -DP_{y}^{-}-V_{x}P_{z}-V_{0}P_{x}^{-}+V_{L}P_{x}^{+}\nonumber \\
\dot{P}_{z}^{-} & = & V_{x}P_{y}^{-}.
\nonumber 
\end{eqnarray}

\begin{figure}
\includegraphics[width=3.25in]{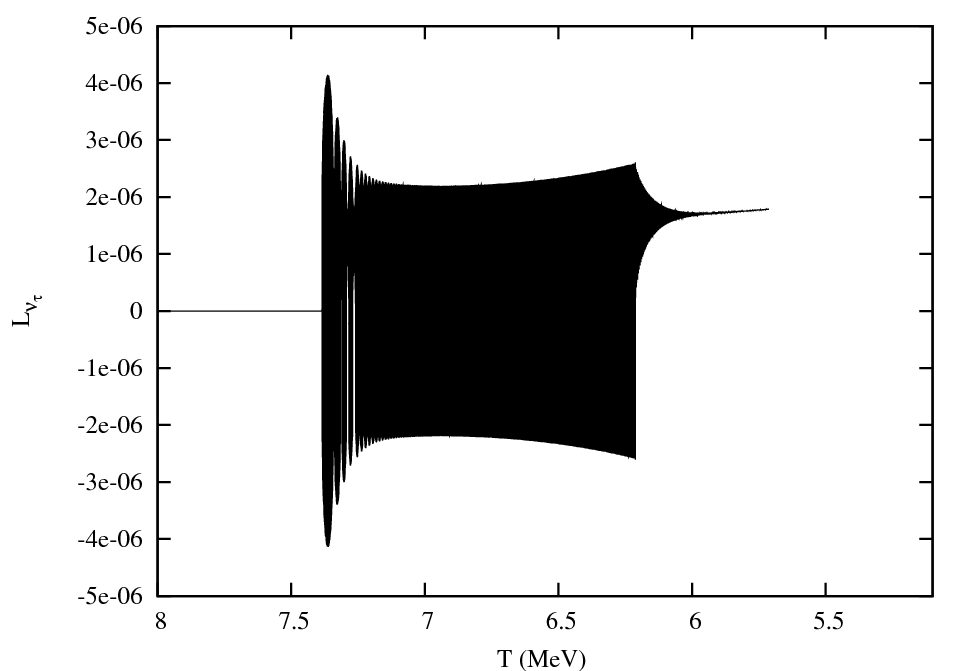}
\caption[]{Shown is the evolution of the lepton number over
  temperature for a particular case of $\nu_\tau\leftrightarrow \nu_s$
  mixing, $\delta m^2 = -0.01\rm\ eV$, $\sin^2 2\theta = 2\times
  10^{-5}$.  \label{1param1}}
\end{figure}

These are coupled non-linear differential equations, with no
analytical solution. The system is seen to undergo resonance at
$V_{0}=0$, if $\delta m^{2}<0$. The asymmetry $L_{\nu_{\tau}}$
undergoes rapid oscillation at the resonance, finally settling down to
a power-law-like growth. See Fig.~\ref{1param1} for an example of the
oscillatory evolution.  The magnitude of the final asymmetry is given
by the solution of $V_{L}+V_{0}=0$, and is given by
$\left|L_{\nu_{\tau}}\right|\simeq11\left(\left|\delta
    m^{2}\right|/{\rm eV}^{2}\right)\left({\rm MeV}/T\right)^{4}$.
After resonance, $L_{\nu_{\tau}}=0$ becomes an unstable extremum, and
the system is trapped into one of the two minima given above.  A
detailed discussion of how the asymmetry evolves can be found in
Ref. \cite{enqvist-1999-464}. The final sign of the asymmetry,
however, is not fixed. It is seen to be very sensitive to initial
conditions, mixing parameters, even choice of step sizes and
integrators \cite{shi-1996-54}.

The differential equations (Eq.~\ref{eq:qke}) are expressed in terms
of time-derivatives but from a cosmological point of view, we are
interested to look at the solutions in terms of temperature. The
time-temperature relation is given by $dt/dT\simeq-M_{\rm pl}/5.44
T^{3}$, $M_{\rm pl}$ being the Planck mass.

\section{Results}
\label{sec:results}

\begin{figure}
\includegraphics[width=3.25in]{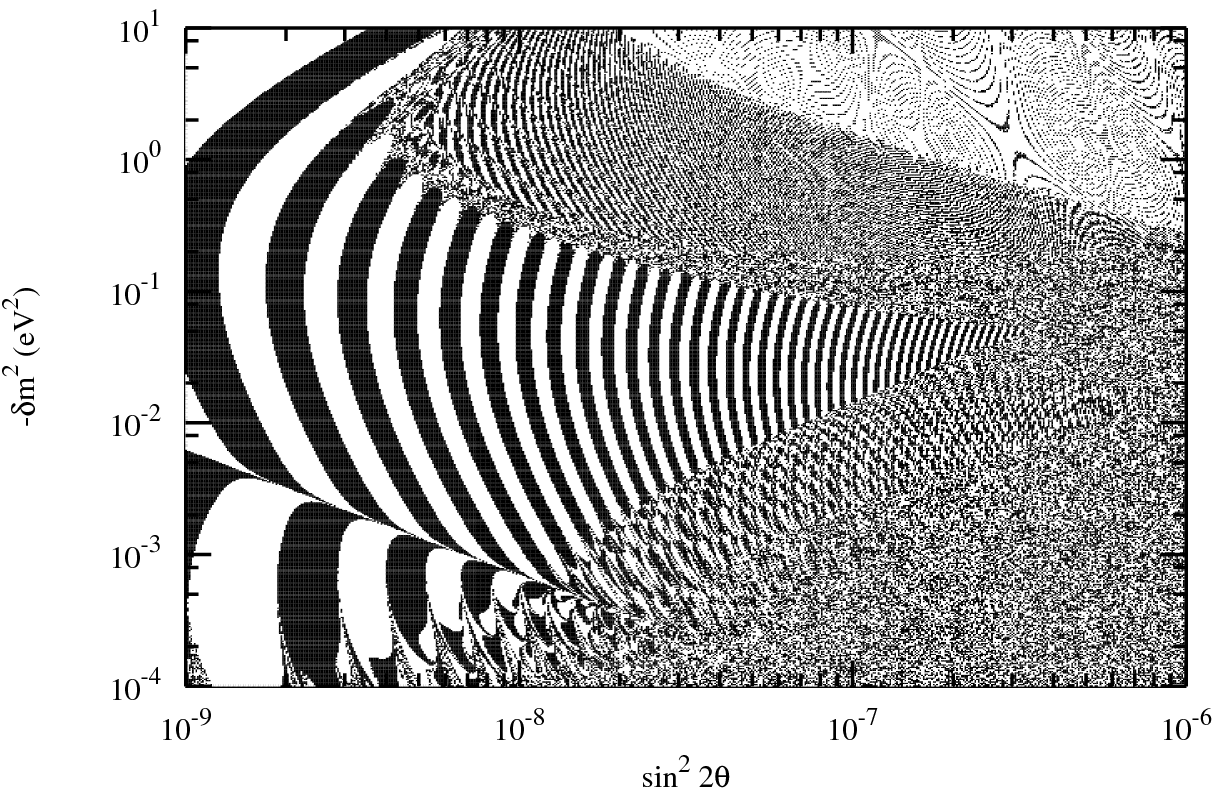}\\
\vskip 0.3cm
\includegraphics[width=3.25in]{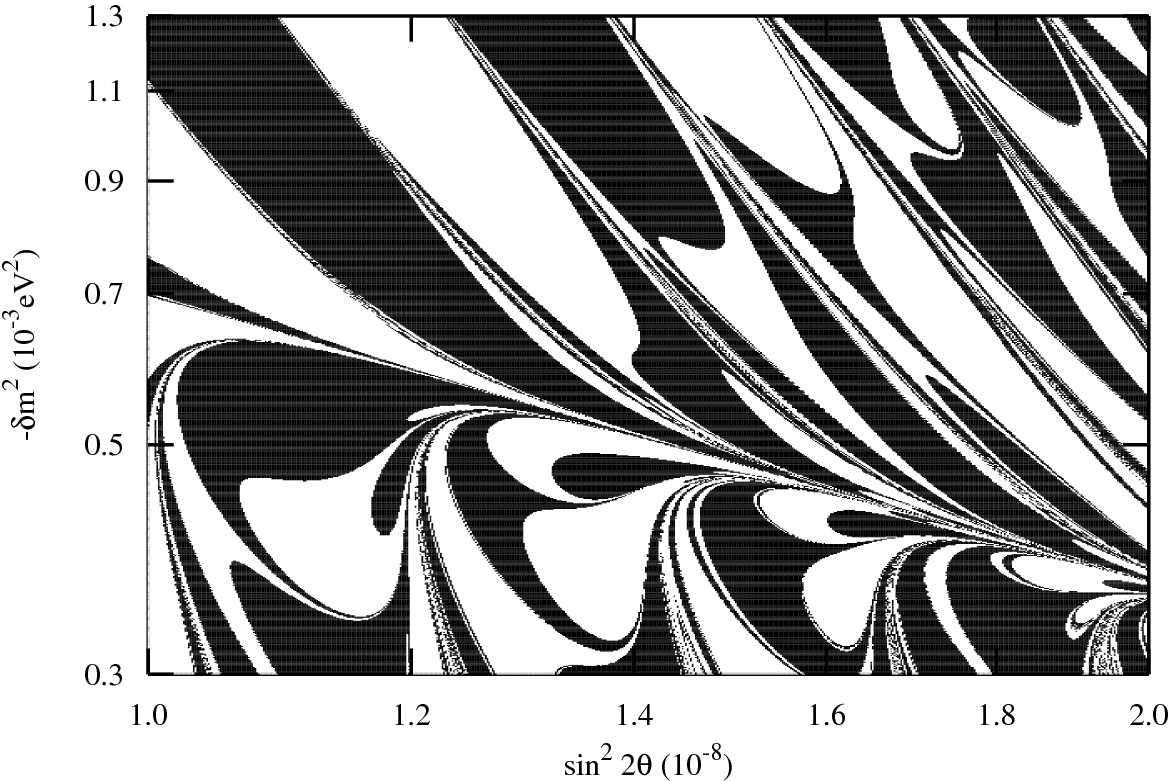}
\caption[]{Shown is the final sign of the lepton asymmetry as a
  function of mass splitting and mixing angle in the case of
  $\nu_\tau\leftrightarrow \nu_s$ mixing for a wide range of parameter
  space (top), and a zoomed region (bottom).  White indicates a
  positive sign and black negative. \label{parameterrange}}
\end{figure}


\begin{figure}
\includegraphics[width=3.25in]{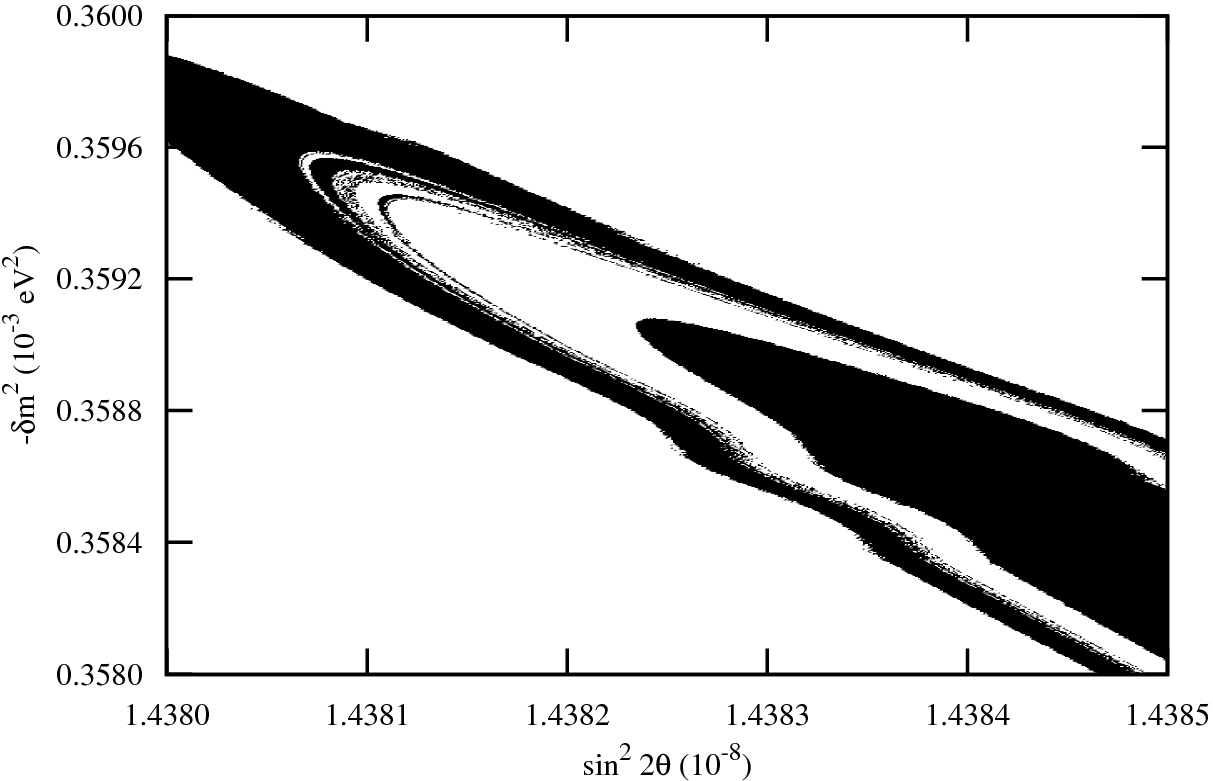}\\
\vskip 0.3cm
\includegraphics[width=3.25in]{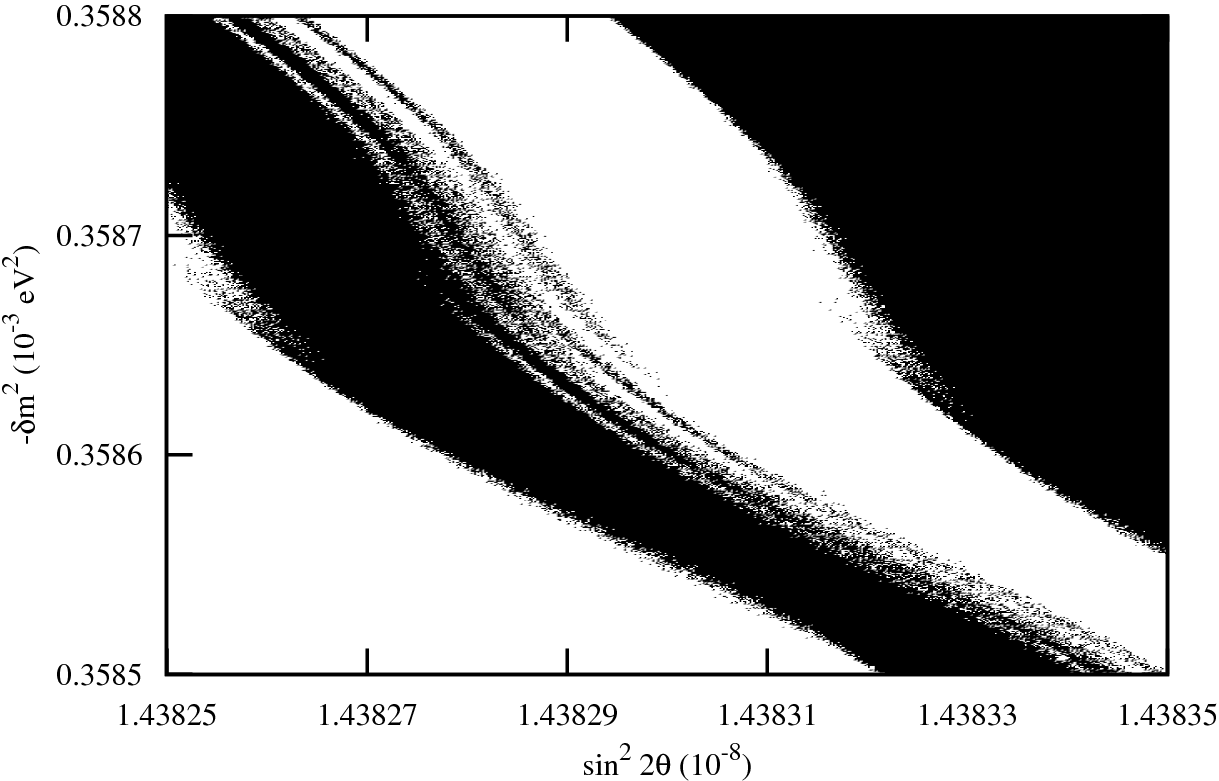}
\caption[]{Shown is the final sign of the lepton asymmetry as a
  function of mass splitting and mixing angle in a 
  zoomed parameter range to show detail of the sign dependence.\label{zoom}}
\end{figure}

We plot the asymmetry as a function of temperature for a couple of
parameter sets, and our results match up well with previous work
\cite{enqvist-1999-464}.  When we scan the entire parameter range
shown in Fig.~\ref{parameterrange}, we note some interesting
facts. First, in our calculations, the band-structure seems to
continue up to many more bands as compared to seen in
Ref. \cite{enqvist-1999-464}.  Further, we can make out structures
even beyond the primary band structure, which has previously been
described as pure white noise. Our numerical routines show that the
number of oscillations increases as the value of $\theta_{0}$
increases, and hence the number of steps for the evolution of the
differential equations increases. This requires that the absolute
error per step in the differential equation be very small in order for
the final sign to be numerically stable. Also, beyond the primary
bands, the structure becomes very fine, and an analysis with a coarse
resolution will just sample the fine structure as an apparent random
sign distribution.

To resolve the fine structure, we zoom into a part of the parameter
space. We see that the region which shows no interesting features in
entire range, shows definite structure when we improve the resolution
in Fig.~\ref{zoom}, with further zoom in to the edge of the bands seen
in figure. The edge is seen to have structure at very high zoom
magnitudes. This behavior is replicated when we zoom into other
parameter ranges. This suggests that the sign of asymmetry exhibits
the characteristics of a fractal with respect to mixing parameters.

We employed the Gnu Scientific Library extensively for our numerical
analysis\footnote{http://www.gnu.org/software/gsl/}.  The system of
differential equations are extremely sensitive numerically, and
therefore demand a very careful treatment. We use the adaptive
Runge-Kutta-Fehlberg routine for our system. For the momentum averaged
solutions, we use a very stringent error bound per step of evolution,
so that the accumulated error does not affect the final sign of
asymmetry. We check that our results converge for a more stringent
error bound. As $\theta_{0}$ and $-\delta m^{2}$ increase, so does the
number of resonant oscillations. This in turn implies that the step
size would decrease in order to follow the resonance correctly, giving
us a larger number of steps overall.  We kept our error bound near
$1\times10^{-15}$, and we found that the error begins to dominate the
solution around $1\times10^{-10}$.  This may be why we see
significantly more structure in Fig.~\ref{parameterrange} than what
was previously seen, in addition to the high resolution sampling of
the parameter space.

To characterize the nature of the fractal solutions, we estimate its
fractal dimension with the correlation integral
\begin{equation}
C(\epsilon) = \lim_{N\rightarrow \infty } \frac{1}{N^2}
\sum_{i,j=1}^{N} \Theta\left(\epsilon - ||\vec{x}_i-\vec{x}_j||\right),
\end{equation}
where $\vec{x}_i$ are the positions of the positive sign lepton number
in parameter space and $\Theta(x)$ is the Heaviside function.  For
finite $N$ the integral is the correlation sum.  For fractal
structures, at high $N$ the sum tends to
\begin{equation}
C(\epsilon) \sim \epsilon^\alpha,
\end{equation}
where $\alpha$ is the correlation dimension, an approximation of the
fractal dimension.  For the range $-0.003{\rm\ eV^2}<\delta m^2 < -0.001
\rm \ eV^2$, $10^{-8} < \sin^2(2\theta) <2\times 10^{-8}$ we find a
power-law dependence of the correlation sum, and arrive at a
convergence of the correlation dimension of approximately 1.05 for
$N \gtrsim 10^4$.  For larger mixing angles, we obtain higher correlation
dimensions of 1.35.

\section{Conclusion\label{sec:Conclusion}}

We have explored the generation of cosmological lepton number in the
case of the presence of a light sterile neutrino state with small
mixing with an active neutrino.  Consistent with previous studies of
lepton number generation using the momentum averaged quantum rate
equations, we find that the process is chaotic and highly sensitive to
the mixing parameters coupling the active and sterile neutrino
states. 

Our main results are the fractal nature of the dependence on the sign of
the lepton number to the mixing parameters as well as the initial
baryon number.  This leaves significant implications for the
predictability of the lepton number of the universe, even in the case
of arbitrarily precise determinations of the mixing parameters between
the active and sterile states in certain regions of the parameter
space.  Future searches for subdominant mixings with the active
neutrinos and unitarity in the active neutrino mixing matrix therefore
could be indeterminant in predicting the sign and magnitude of the
cosmological lepton number.

The fractal or correlation dimension depends on the region in
parameter space that is examined, which is expected.  The banded
structure of the sign of the parameter space undergoes changes in its
structure throughout the space, with some definitive lines in the
pattern change.  The fractal dimension therefore shifts in the
parameter space.  This is likely due to the fact that the frequency of
the rapid oscillations are deterimined by the oscillation dynamics,
and the point at which they stop or freeze-out is determined by the
expansion dynamics, which depend on the total radiation energy density of
the universe independent of the neutrino physics.  

Our work has focused on the single-momentum quantum rate equations for
the solution of this system, and we acknowledge that it has been shown
that the behavior changes in the full momentum-dependent quantum
kinetic description.  Studies have found that the oscillatory behavior
persists and may be chaotic in the full quantum kinetic equation
solution \cite{bari-foot,kainulainen-2002-0202}.  Since there are
numerical difficulties in quantifying the nature of the solution in
the full momentum case, which we find in our own work, the resolution
of this question may require a completely new formulation of the
quantum kinetic equations and their solution in order to make them
numerically stable to provide a convincing solution of the evolution.
The work presented here highlights the new type of chaotic, fractal
physics that can arise in this system, and its consequences in
determining the lepton number of the universe given the existence of a
light sterile neutrino.  If a light sterile neutrino is discovered
within this parameter space, a new approaches to quantifying the
lepton number of the universe will be strongly motivated. Though if
the sign of the lepton produces such fractal structure, the sign of
the lepton asymmetry may be indeterminate due to arbitrarily fine
fractal structures.

\begin{acknowledgments}
  We thank John Beacom and Nicole Bell for discussions in very early
  stages of this work.  We also thank Edward Ott for useful
  discussions.  This work was supported at the University of Maryland
  by the Maryland Center for Fundamental Physics.
\end{acknowledgments}
\appendix

\bibliography{ref}

\end{document}